# Development of Single-Shot Multi-Frame Imaging of Cylindrical Shock Waves in a Multi-Layered Assembly

Leora Dresselhaus-Cooper,[a,b] Joshua E. Gorfain,[c] Chris T. Key,[c] Benjamin K. Ofori-Okai,[a,d] Suzanne J. Ali,[e] Dmitro J. Martynowych,[a,b] Arianna Gleason,[d,f] Steven Kooi,[b] Keith A. Nelson[a,b]

[a] *Department of Chemistry, Massachusetts Institute of Technology,*

[b] *Institute for Soldier Nanotechnology, Massachusetts Institute of Technology*

[c] *Applied Physical Sciences,* [d] *SLAC National Accelerator,* [e] *Lawrence Livermore National Laboratory,* [f] *Los Alamos National Laboratory*

**Abstract**

We demonstrate single-shot multi-frame imaging of quasi-2D cylindrically converging shock waves as they propagate through a multi-layer target sample assembly. We visualize the shock with sequences of up to 16 images, using a Fabry-Perot cavity to generate a pulse train that can be used in various imaging configurations. We employ multi-frame shadowgraph and dark-field imaging to measure the amplitude and phase of the light transmitted through the shocked target. Single-shot multi-frame imaging tracks geometric distortion and additional features in our images that were not previously resolvable in this experimental geometry. Analysis of our images, in combination with simulations, shows that the additional image features are formed by a coupled wave structure resulting from interface effects in our targets. This technique presents a new capability for tabletop imaging of shock waves that can be easily extended to large-scale facilities.

**Introduction**

The destructive and variable nature of shock waves places high importance on techniques that can provide time-dependent observations in a single experiment.[1,2] Multi-frame imaging of shock waves generated by projectile or explosive impact has been demonstrated on length scales of microns through many meters and time scales of nanoseconds through seconds.[3,4] Laser-generated shock waves are typically monitored optically, including by imaging; however in most cases the shock wave propagates toward the observer.[1,5] The temporal resolution of measurements that image the shock is set by the relationship between the shock velocity and the imaging resolution.[6,7] Imaging methods can also record spatial variation in the shock profile by measuring the side-on spatial profile of the shock.[7] Viewing the waves perpendicular to their direction of propagation allows for monitoring of the spatiotemporal evolution of shock waves as they move through a material.

Pezeril *et. al.*[6] developed a method for laser generation of converging shock waves in a layered sample assembly consisting of a thin (~ 10 μm) liquid layer sandwiched between two glass substrates. The shock wave was launched by a laser pulse that was focused to a 200-μm diameter circular "ring" pattern at the sample. Absorption of the light in the liquid layer (India ink, i.e. water with amorphous carbon) initiated quasi-2D shock propagation and focusing which increased the pressure of the laser-induced shocks by an order of magnitude, enabling this tabletop experiment to reach high pressures that are usually only attainable in flyer plate, gas gun, or high energy laser facilities.[8,9] This table-top approach enables us to develop new diagnostic techniques to study converging shock waves and the high-pressure dynamics they induce. The geometric instability of converging shock waves makes it difficult to track their propagation with conventional diagnostics that disrupt the wave geometry and thus its properties.[8] As experiments on converging shock waves typically require specialized ultrahigh-energy laser facilities,[8,9] detonation chambers,[10] or high-current facilities,[11] diagnostic development is challenging because of the limited facility time allocated to each experiment. In our case shock propagation in the sample plane, perpendicular to the direction of the light beams, can be monitored by imaging with simultaneously high spatial and temporal resolution. The shock profile can be extracted using interferometric imaging and an empirical formula relating the refractive index to the density.[12] This previous work gathered sequences of images showing the shock's trajectory by assembling single-frame measurements with varied excitation and imaging pulse delays, acquired at different regions of the sample since each shocked region was permanently altered.

While this technique proved informative for studying low-pressure, reproducible shock waves in samples that were essentially uniform spatially, challenges remained in characterizing the dynamics of events that differ from shot to shot, such as geometric instabilities near the center of convergence,[13] fracture,[14] and shock-induced decomposition of energetic materials.[15] To measure these and other phenomena whose details are not reproducible, we need to collect the entire image sequence in a single shock event.

In this paper, we present an extension of the capabilities of our previously reported technique[6,12,16] by using a multi-frame imaging method to record dynamical information in a single shot. A Fabry-Perot cavity produces a sequence of ultrafast imaging pulses with a temporal separation that is adjusted to match the timing of a sequence of electronically gated CCDs in a high-frame-rate multi-frame camera. We demonstrate the utility of this multi-frame imaging technique in both shadowgraph and dark-field modes to acquire image sequences showing the progression of the shock wave through our targets. The image sequences capture geometrical distortions in the wave and high-order interface interactions that vary on a shot-to-shot basis due to fluctuations in the drive beam and small variations in the sample. These image sequences show multiple converging and diverging shock features which we compared to hydrodynamic simulations, revealing a coupled-wave structure caused by the interfaces in the targets. Thus single-shot multi-frame imaging has revealed important features that single-frame images could not discern about shock propagation in our multi-layer targets and (including geometric instability as well as coupled-wave effects at interfaces) that have rarely been observed directly in any sample geometry.

**Experimental Methods**

The experimental setup is illustrated in Fig. 1. We used a multi-stage amplified Ti:Sapphire laser system, with the oscillator output stretched to 150 ps and amplified at a 10 Hz repetition rate to 4 mJ in a regenerative amplifier and to 30 mJ in a six-pass bowtie amplifier. We used a half-wave plate and polarizer as a variable beamsplitter to separate 4 mJ of the amplified uncompressed pulse for use as the drive pulse. The rest was compressed to 130 fs and used for the imaging probe.

The drive pulse was shaped into a 150 μm inner diameter ring (8 μm ring width) by passing the 800 nm uncompressed pulse through a 0.5° axicon (a conical prism) and an $f = 30$ mm lens. The femtosecond probe pulse was passed into a frequency-doubling Fabry-Perot cavity, and the output sequence of probe pulses was directed through the target and imaged onto single- and multi-frame cameras. We positioned the sample within the excitation ring pattern using the single-frame camera. The Specialized Imaging SIM 16X collected up to 16 images with as short as 3 ns inter-frame intervals on electronically gated intensified CCDs.

The target assembly had a ~10-µm thick layer of 10 vol% India Ink (an aqueous solution of amorphous carbon) in deionized water placed between two 250 µm thick, 25.4 mm diameter *r*-cut sapphire wafers. Interaction of the intense drive light by the amorphous carbon in the ink launched high-amplitude expansion waves that formed a shock.[6] All the experiments were conducted using a drive laser pulse energy of 1.0 mJ, corresponding to a fluence of 26.5 J/cm$^2$ and an intensity of $1.70\times10^{11}$ W/cm$^2$, to generate the shock waves. Within the sample plane, expansion from the dye combustion formed two shock waves in water – one diverging outwards, and the other converging inwards toward the center of the ring. Additional features are described in the Results and Simulations sections.

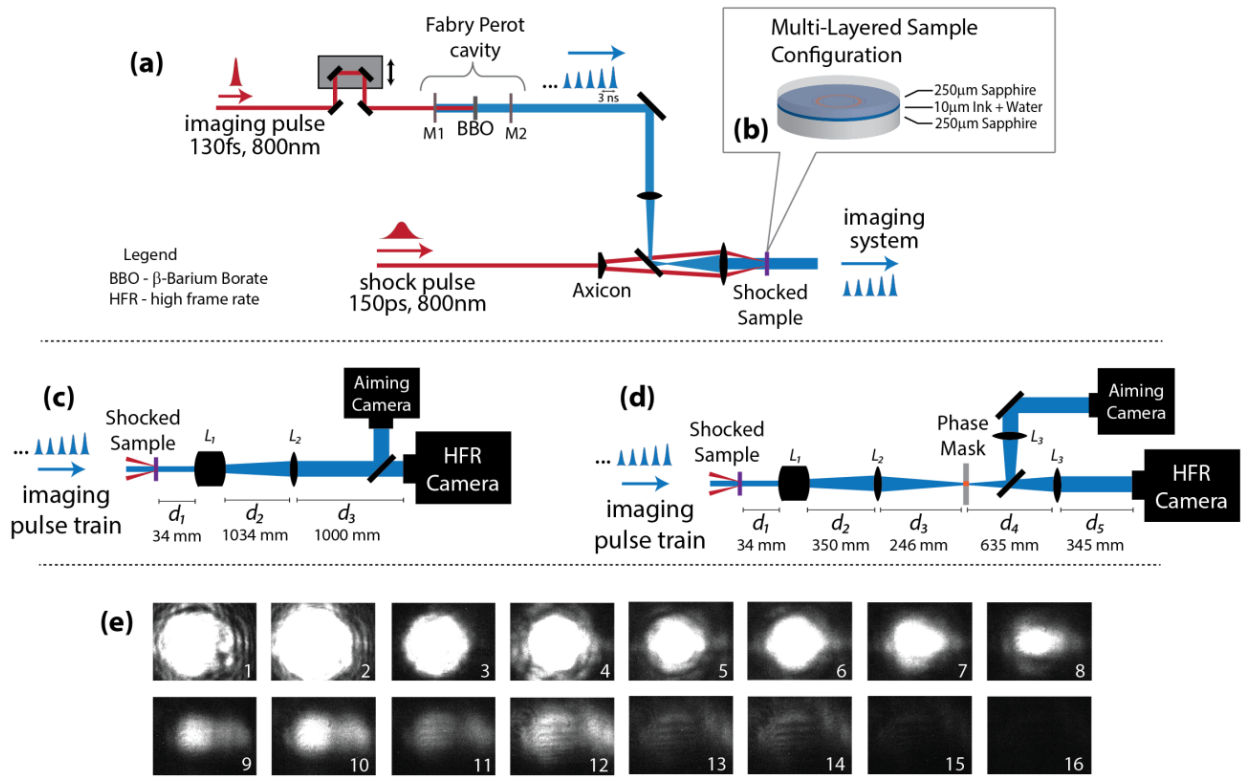

**Figure 1.** Schematic representations of (**a**) the optical configuration for shock generation, with a Fabry-Perot cavity to create a train of ultrafast imaging pulses for multi-frame imaging diagnostics. (**b**) The multi-layered target configuration used for this work, with a depiction of the drive laser excitation "ring" pattern and expected shock behavior. The optical configurations for (**c**) shadowgraph, and (**d**) dark-field imaging of the shocked sample. (**e**) Image sequence showing the decreasing illumination of each successive imaging pulse output from the Fabry-Perot cavity. The pulses were apertured with an iris after the cavity, and detected with CCDs using no electronic gain. All subsequent image sequences use the detector gain to standardize the image brightness between frames.

The essential components of our imaging setups are shown schematically in Figure 1. For each imaging system, ~10 mJ of compressed light was directed from the laser into the probe beam. The probe was passed through an adjustable optical delay and then into a doubling Fabry-Perot cavity to generate a pulse train (Fig. 1a). The cavity consisted of a pair of mirrors, M1 and M2, which were highly transmissive ($T_{800} > 99\%$) at 800 nm wavelength, but reflective to 400 nm wavelength ($R_{400} > 99\%$ for M1 and $R_{400} = 93\%$ for M2). After coupling into the cavity through M1, some of the 800 nm pulse was frequency-doubled to 400 nm light using a β-barium borate (BBO) crystal. The unconverted 800 nm light exited the cavity through M2 without reflecting, and was removed with a reflective filter (OD > 6 at 800 nm) after the cavity. As mirror M2 was 7% transmissive to 400-nm light, and the BBO crystal had 6% losses for each pass through the crystal (reflection and absorption), 82% remained in the cavity after each successive round trip (87% after the first pass). Frequency-doubling and reflective losses within the cavity generated the pulse train with 21% conversion efficiency (with 5.75 mJ of 800-nm input) for the pulse train leaving the cavity. Each frame was normalized in intensity by adjusting the gain at the corresponding CCD and by post-processing to adjust the white balance. Random noise from high gain was compensated for with image processing by smoothing with Speckle filters, as described in the Supplemental Information. The time delay between successive pulses was determined by the cavity length, and measured on a digital oscilloscope. The probe spot was telescoped to a 1.5 mm diameter waist at the sample.

Two separate imaging systems—shadowgraph and dark-field—were used to measure the shock progress in the plane orthogonal to the beam propagation direction. In shadowgraph imaging (Fig. 1c), maps of the amplitude of the light passing through the sample were collected with a narrow depth of focus (~4 μm) within the image plane. For this system, $L_1$ was a 10X infinity-corrected objective, used to acquire good spatial resolution and image quality, and $L_2$ had $f = 1000$ mm to expand the field of view. Dark-field imaging (Fig. 1d) created spatial maps showing variations in the refractive index resulting from the shock. The system was constructed using a three-lens imaging system, with $L_1$ the same 10X objective, $L_2$ an $f = 150$ mm lens, and $L_3$ an $f = 300$ mm lens. Phase-to-amplitude conversion between the sample and detectors was achieved by placing a mask in the Fourier plane of the object within the imaging system to block the $0^{th}$-order light that was not deflected significantly from its original wavevector direction after passing through the sample assembly. The mask was an optically opaque gold rectangle of 30 μm x 60 μm x 10 nm dimensions on the surface of a 2 mm thick 25 mm x 25 mm fused silica double-sided optical flat.

Using the imaging systems, we resolved features < 2 μm in size with a field of view of 131 μm x 171 μm for the shadowgraphs and 332 μm x 442 μm for the dark-field images. The 3 ns duration between frames was the shortest possible in this configuration, as set by the capabilities of the camera.

**Experimental Results**

*I. Shadowgraph imaging*

Figure 2 shows a sequence of shadowgraph images that demonstrate our single-shot method to image cylindrically propagating shock waves. Image sequences like the one shown here allow us to track non-reproducible events like the geometric instability seen in Figure 2 and discussed later in this section. Shadowgraph image signal comes from probe light refraction due to refractive-index changes in the material, in this case caused by density variations from the shock wave.[5,17] The imaging light is sensitive to the spatial second derivative of the refractive index, $\nabla^2 n = \frac{\partial^2 n}{\partial x^2} + \frac{\partial^2 n}{\partial y^2}$, in the object plane.[5] This means that our shadowgraph image sequences are most sensitive to the shock front or other abrupt index changes in the target.

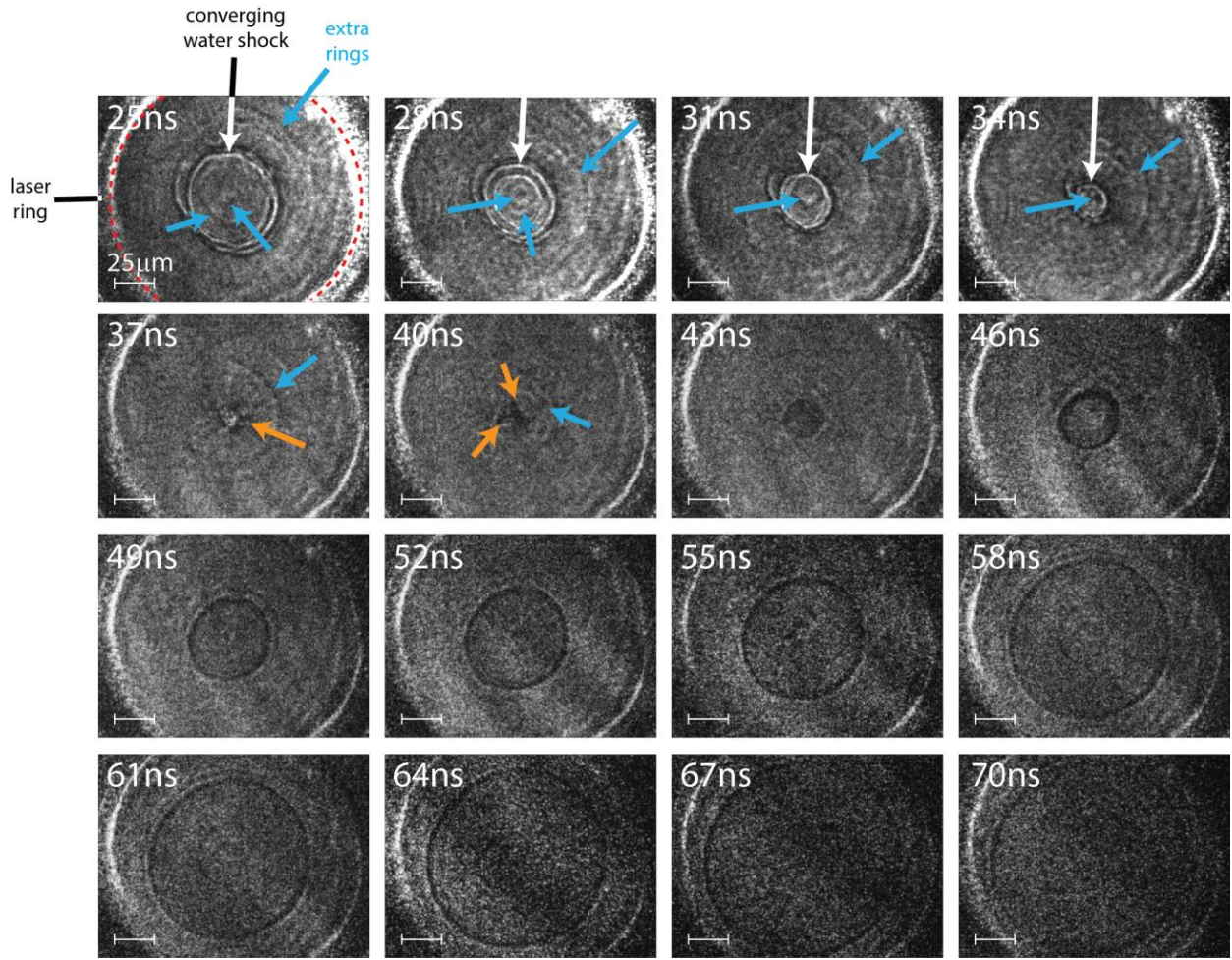

**Figure 2.** Single-shot multi-frame shadowgraph images showing shock convergence and subsequent divergence in the multi-layered target. The inner edge of the laser excitation line is shown in red in the first frame, white arrows point to the water shock front, blue arrows show additional image features, and orange arrows point to geometric instabilities. The image at 28 ns includes a ghost image due to some light from the 31-nm probe pulse reaching the 28-ns CCD.

The 16 frames in Figure 2 are separated by 3 ns intervals, with the 130-fs probe pulse duration setting the integration time for each image. The laser excitation ring, highlighted with red dashed lines in the first frame, produced the white outer ring that is stationary throughout the entire sequence. Frames from 25 ns to 37 ns display concentric rings that get smaller with increasing time, corresponding to the shock front in water converging to the center of the excitation ring. The 40-ns frame shows the dark feature at the center of convergence, with slight geometric distortions, and all subsequent images (43-70 ns) see the rings expanding outward as the shock diverges. We observe imperfect circles from the converging shock front (25-37 ns), indicating geometric instability that originates from inhomogeneities in the mode of the drive laser.[18,19] Other image sequences also show instability, but the precise evolution is different in each shock event. While the converging wave's geometry distorts as it approaches the focus, the diverging wave maintains a nearly circular structure. This type of geometric instability is characteristic of converging shock waves,[20–22] but its details near the center of convergence have rarely been resolved as they are by the short temporal and spatial intervals between successive images in our measurements.

Close examination of all frames of the image sequence shows faint concentric rings, shown with blue arrows, in addition to the main ring. These features are most clearly seen in the frames collected at 25 ns, 31 ns and 37 ns, although they are present in every frame. As we will describe later on, tracking these faint image features enables us to follow shock behavior (e.g. substrate shocks, coupled wave interactions, etc.) that would be difficult to discern in single-frame experiments.

## II. Dark-field imaging

Dark-field imaging enabled us to gather a complementary multi-frame view of shock progression in our multi-layered targets. In dark-field imaging, the signal originates from variation in the optical phase that the probe beam accumulates as it propagates through the transparent target. As the imaging light field propagates through the target, it acquires a phase $\phi = \frac{2\pi n}{\lambda}\ell$, where $n$ is the refractive index, $\ell$ is the thickness of the target in the region with that value of refractive index, and $\lambda$ is the wavelength in air. These cylindrical shock waves change the refractive index along the radial coordinate (distance from the center of convergence) $R_S$, the angular coordinate $\theta$, and the target depth, $Z$. Consequently, the total accumulated phase is $\phi(R_s,\theta) = \phi_0 + \Delta\phi(R_s,\theta) = \frac{2\pi n_0}{\lambda}\ell + \frac{2\pi}{\lambda}\int_0^\ell \Delta n(R_s,\theta,Z)dZ$, where $\phi_0$ and $n_0$ are the average phase and refractive index. Density variation from the shock wave creates local refractive-index changes, $\Delta n(R_s,\theta,Z)$, that result in corresponding local phase shifts, $\Delta\phi(R_S,\theta)$, in the imaging light. The value of $\Delta n(R_s,\theta,Z)$ at a given position corresponds to the difference in refractive index between that point and the average index for the $R_S$, $\theta$ plane at that $Z$ position.

We convert the object phase pattern into an amplitude pattern that our camera can detect by placing a phase mask at the focus of $L_2$ (Fig. 1d), i.e. the Fourier plane of the target. Our transformed image appears in the Fourier plane as a bright central peak surrounded by dim deflected light. The bright signal is from imaging light whose phase was not changed significantly from the average, e.g. light that encountered unshocked material, while the dim surrounding features come from phase shifts produced by the shock wave.[23,24] By blocking the undeflected light with the gold mask, we cause the light hitting our detector to originate only from the deflected, phase-shifted light that was influenced by the shock.

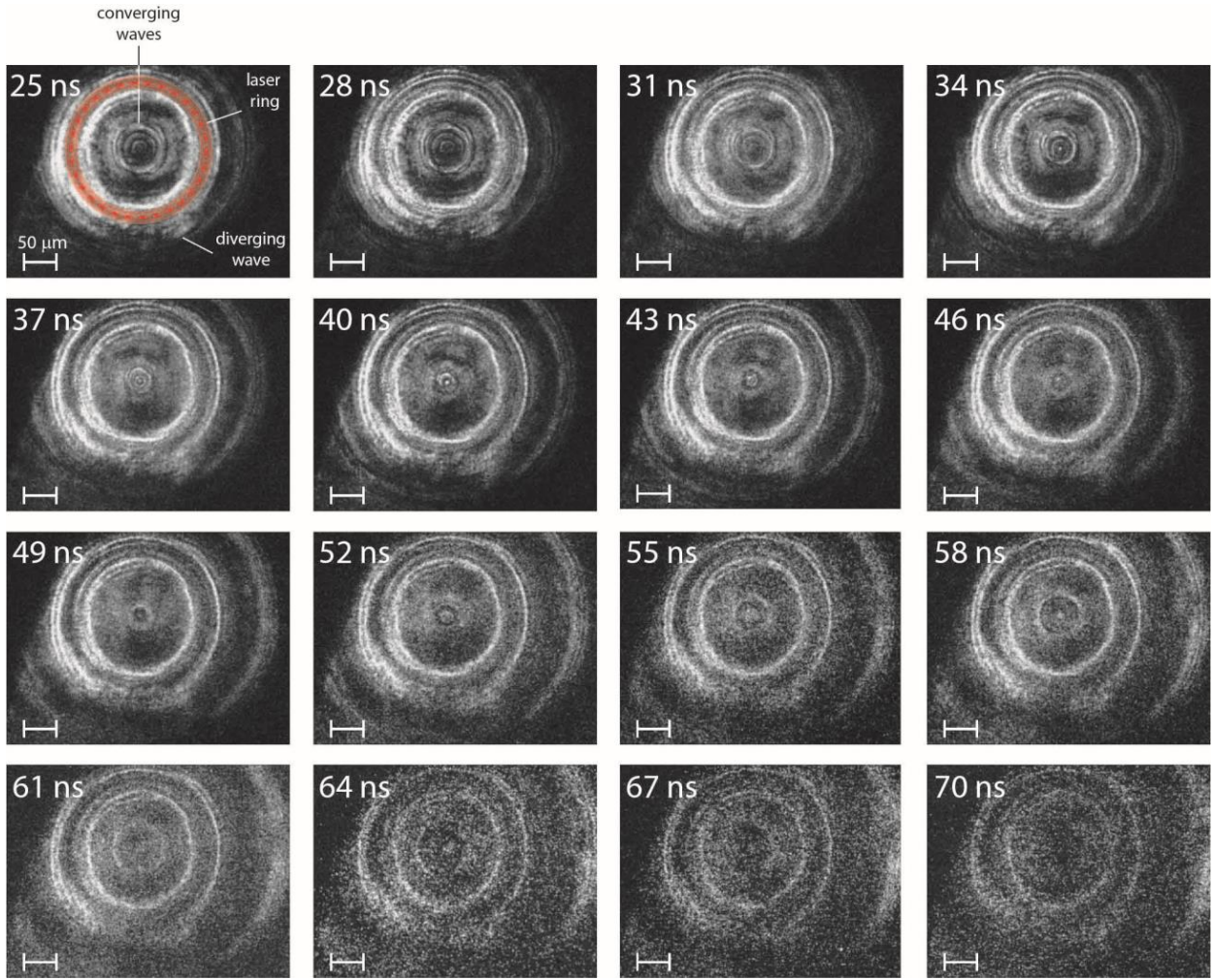

**Figure 3.** Dark-field image sequence showing shock convergence and subsequent divergence in the multilayered target geometry. The shock was generated with the same drive conditions as in Figure 2. The bubbles from the drive laser excitation ring are indicated by the red line and translucent red band in the first frame.

We show a dark-field image sequence collected with the multi-frame configuration in Figure 3. The intensity of light at each pixel is proportional to 1–cos($\Delta\phi$), revealing regions where there were large phase shifts accumulated through the target. Bright features originate from changes in the refractive index, which are dominated by changes to the density.[24] Signal from bubbles produced by the laser excitation ring are highlighted by a translucent red band and a red dashed line in the first frame, and this signal is repeated in each subsequent frame. The images capture a larger field of view (lower magnification) than those in Figure 2, showing the diverging shock outside the laser excitation ring in addition to the converging shock within.

Consistent with the shadowgraph sequence, we observe multiple concentric rings inside the laser ring for each image across the entire 25-70 ns period captured in the sequence. Both image sequences enable us to track multiple rings, but shot-to-shot experimental variation prevents precise correspondence between different sequences. While multiple rings are quite clear for the waves inside the laser excitation ring, they are not discernable for the lower amplitude diverging wave outside the excitation ring. The unusual behavior of these extra rings suggests that their cause is strongly dependent on the shock amplitude, which is highest for the converging wave inside the laser excitation ring. Multi-wave structure in shocks can be caused by phase transitions in the material,[25] but as discussed further below, in this case it is more likely a result of only partial confinement of the shock in the water layer.

**Simulation Results**

The simulations were conducted using the CTH shock physics code, developed at Sandia National Laboratories,[26] with setup parameters matching those from the experiment as closely as possible. We used the simulated density maps to calculate the refractive index variations in the target and the phase shifts imparted to our imaging light in order to model the images we observed. To understand the origin of the additional rings we observed in both image sequences, we extended previous simulations[6] to include all five target layers: Air, Sapphire, Water, Sapphire, Air. Due to experimental uncertainties in the water layer thickness (10-20 μm) and quantitative differences between the real and simulated shocked sample assemblies (for example, the simulation does not account for sapphire anisotropy), the simulations provide a qualitative picture of shock propagation in the targets. Pressure-dependent values of the water photoelastic constants[27–29] were used to calculate the shock-induced refractive index changes in the water sample layer.

Figure 4 shows an axisymmetric view of the samples, with the color-code indicating the pressure at each radius $R_S$ and depth $Z$ in the target. Figure 4a shows the result of the initial heating of the irradiated water region, causing high pressure and thermal expansion which launches shock responses in all directions. Expansion in the vertical (Z) direction launches shocks that propagate and diverge in the sapphire substrates. These are clearly observed in Figures 4b-d. Some components of these waves reach the sapphire-air interfaces (see Fig. 4e) and later (at times t > 50 ns) return to the water layer, but by then they have diverged and have negligible pressure. Thermal expansion of the irradiated water region in the sample plane launches the water shock that is cylindrically focused and reaches the highest pressure at the center of convergence as seen in Figs. 4b-d. (The diverging water shock is also apparent in Figs. 4b-d, moving away from the focal region.) We will refer to this quasi-confined water wave as the primary shock. The inset at 22 ns (Fig. 4d) shows the highest pressure reached in the simulation, 21 GPa, in the water layer at the center (observed in our shadowgraph images at 40 ns).

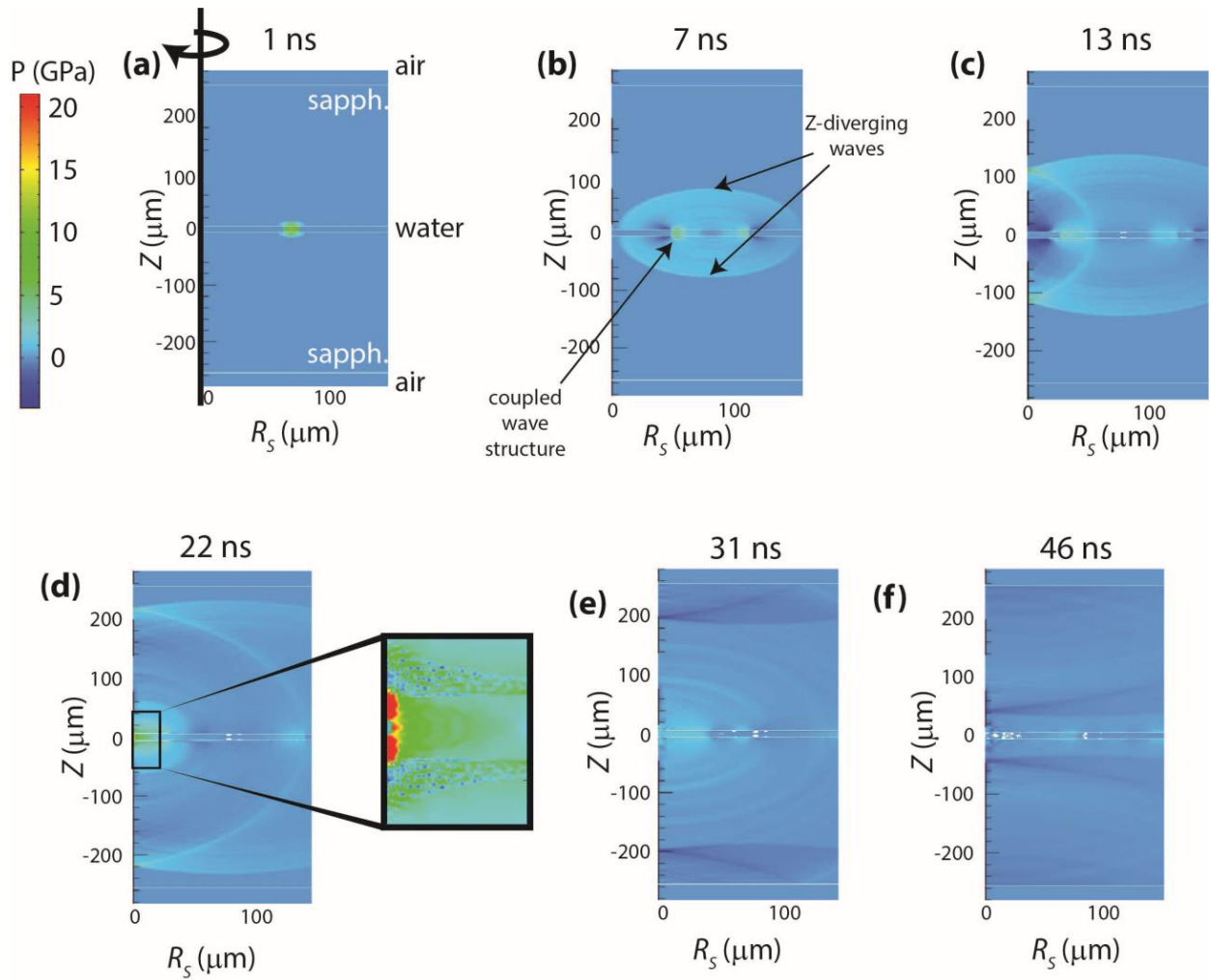

**Figure 4.** Axisymmetric view of the shock simulated across the target geometry, displaying pressure changes along the depth and radius. The time steps show the shock convergence and subsequent divergence, and indicate the Z-diverging and coupled-wave shock components.

Because the water-sapphire impedance mismatch leads to only partial confinement of shock pressure within the water layer, during its convergence the primary shock transmits part of its mechanical energy into the substrates, launching the hemi-torroidal shock waves in sapphire that converge along $R_S$, while diverging in $Z$.

The primary water shock also extends through the interfaces and into the nearby regions of the substrates as it propagates toward the focus. The shock speed is faster in the substrates, and the shock running ahead in the substrates also extends back into the sample layer. This coupled-wave structure consists of the primary water shock, an oblique sapphire shock in both of the nearby substrates, and an oblique water pre-shock. The primary water shock leaks an oblique shock of lower energy into the adjacent sapphire substrates, which travels fastest because of sapphire's higher impedance. The sapphire waves leak additional weak oblique waves back into the water layer, which trail the sapphire waves due to water's lower impedance.

**Discussion**

We now consider the manner in which the simulated shock structure gives rise to distinct features in imaging measurements, and we compare the results to the experimentally measured features including multiple rings. The coupled-wave structure extends through the water layer and into the nearby substrate regions, but water's far larger photoelastic constant means that the images we observe are primarily due to pressure-induced refractive index changes in the water. Using the hydrodynamically simulated densities and the reported pressure-dependent photoelastic constants[27,28,30], we calculated the predicted refractive index values in each layer. When the primary water wave reaches the center of convergence, the water index varies by $\Delta n_{water} \sim 0.5$, while sapphire's lower shock pressures and photoelastic constants cause each substrate in the immediate vicinity of the water layer to undergo a maximum change of $\Delta n_{sapph} \sim 0.03$. This suggests that the additional rings in our images originate from the water component of the coupled-wave structure. In fact, the sapphire shocks only achieve measurable $\Delta n$ and $\Delta \phi$ values at their highest pressures, near the center of convergence.

The imaging methods capture different aspects of the shock dynamics in our targets. Shadowgraph imaging is sensitive to abrupt and large variations in the refractive index, showing the shock front but not the entire wave.[5] By contrast, dark-field images can capture lower-amplitude and more gradual density changes in the shock because of the signal's accumulated $\mathit{\Delta} n$-dependence.[5,17,31,32] These different signal dependences enable a clear view of the leading oblique waves with dark-field imaging, while resolving the position of the primary water shock front in shadowgraph images.

Dark-field images produce signal from the accumulated phase change experienced by the light passing through the target, making its dependence on refractive index $1 - \cos(\Delta\phi) = 1 - \cos\left(\frac{2\pi\Delta n}{\lambda}\ell\right)$ for each region $\ell$ over which there is little variation in $\Delta n$. Standard dark-field images show a $I \propto \Delta\phi^2$ dependence for small phase shifts, following the small-angle approximation. In our case, the accumulated phase cycles repeatedly over intervals of $2\pi$. Using our calculated refractive index values from the simulation for a 13 ns delay (Fig. 4c), we show the accumulated phase through the water (blue) and sapphire (red) layers of the target in Figure 5b. We associate this simulated delay approximately with the experimental image recorded at 25 ns delay based on the difference between simulated and experimentally measured shock speeds. The plots in Figure 5b shows that the dark-field image includes phase shifts of up to ~$8\pi$ through the water layer, while only ~$0.2\pi$ through the sapphire. The large scale for accumulated phase in Figure 5b suggests that sapphire's phase is constant, but it does vary through the $0.2\pi$ range. The small phase shift in sapphire causes its signal to produce a $I \sim \Delta\phi^2$ dependence, following the small-angle approximation. Firstly, this means that the imaging signal primarily originates from the water layer, though there is a small contribution from each sapphire substrate, as seen in Figure 5c. Additionally, the sinusoidal relationship between image signal and phase cause the water layer to over-rotate the phase shift, creating artificial features in the images from the phase cycling, as seen at 35 μm, 45 μm and >60 μm in Figure 5c. While phase cycling generates extra features, this technique's high $\Delta n$-sensitivity causes density variation within the profile of the coupled-wave structure to be well-resolved in the image signal. The predicted image signal shown in Figure 5c shows that the leading oblique wave should be well resolved in our dark-field images, corresponding to the dark and bright rings shown in Figure 5a. While over-rotation prevents dark-field images from providing a complete view of the coupled-wave structure, its sensitivity gives detail in the structure of the oblique shock front, providing complementary information to the shadowgraph images.

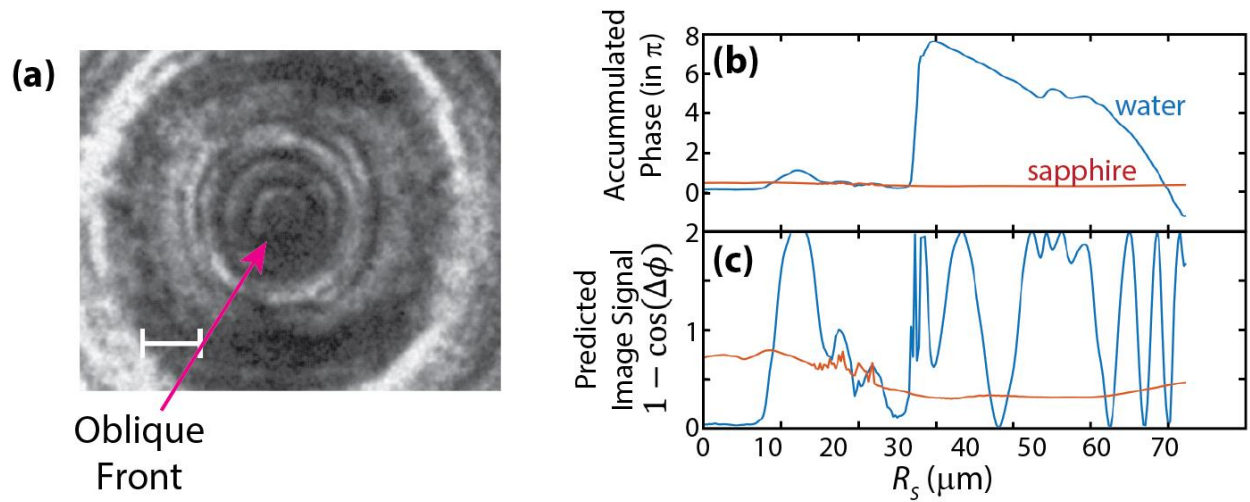

**Figure 5.** (a) Dark-field image collected 25 ns after the drive pulse, compared to the simulated values (from the corresponding 13-ns delay) showing (b) the accumulated phase the light acquires through the sample in units of $\frac{\Delta\phi}{\pi}$, and (c) the predicted image signal $1 - \cos(\Delta\phi)$ showing the result of the sensitivity and the over-rotation.

Our shadowgraph images show significantly fewer concentric rings, with one clear primary ring between two faint ones. Because the signal depends nonlinearly on the variation in $n$, the shadowgraph images produce fewer features. Figure 6b shows the predicted refractive index and its spatial second derivative, $\nabla^2 n$, giving assignments for each ring in our images. We calculated the spatial second derivative using $\frac{\partial^2 n}{\partial x^2} + \frac{\partial^2 n}{\partial y^2}$ on the Cartesian unwrapped image for the 4-μm thick water layer at $Z = 0$, and present a radial slice from the resulting symmetric derivative map in Figure 6c. At $t = 25$ ns in our experiment (corresponding to 13 ns in the simulation), we see one primary ring, with three additional faint rings. Comparison between the simulations and experimental images clearly indicates that the single darkest ring corresponds to the shock front of the primary water wave, which has the highest $\nabla^2 n$ value. Within the image sequences in Figure 2, we can resolve the progression of two additional waves, one preceding and one following the primary shock front. From our simulations, we interpret that both the leading (yellow) and trailing (blue) faint rings originate from the water components of the coupled-wave structure. We attribute the leading faint ring to the oblique water shock, which both the image and predictions show precedes the most intense signal from the primary water wave (Fig. 6b).

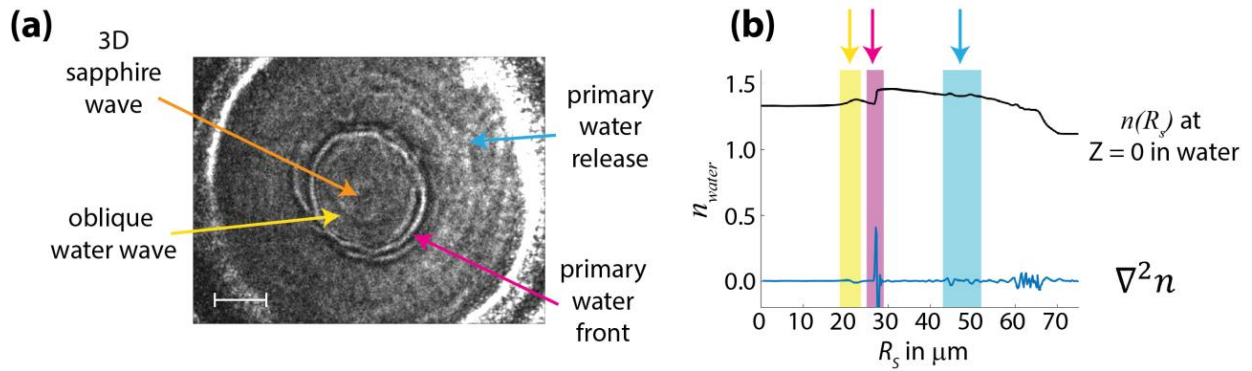

**Figure 6.** (a) Assignments showing the origin of each ring in the shadowgraph image from 25 ns. This is compared to (b) plot showing the simulated refractive index and spatial second derivative (Laplacian) of the refractive index for the center of the water layer as a function of radius. The simulation is from 13 ns, as the simulated shock velocities were faster than those observed.

At the end of the coupled-wave structure, the simulations predict two types of pressure release: a release from the primary water wave, and a subsequent release from the oblique waves. In both cases, the magnitudes of the density changes should be relatively large (largest for the primary water wave), but the release profiles should be relatively slow, causing their second derivatives to be relatively small. This translates to faint signals in our shadowgraph images, as shown by the blue arrow in Figure 6a. Our predictions show that the primary water release (shown in blue) trails the primary feature in subsequent images throughout Figure 2.

The final wave highlighted in the shadowgraph image in Figure 6a does not originate from our view of the water layer, implying that it comes from the sapphire layers. Sapphire's low photoelastic constant causes its refractive index to change by only ~0.08 through each substrate at this simulated 13 ns time, but two factors cause the wave to be visible. In Figure 2, some of the blue arrows point to sapphire rings at the center of convergence in nearly the same position from 25-37 ns in our sequence, which vary in intensity but not position. Between the interface and 3D waves, each substrate sees a wave at the focus at some depth and amplitude over the entire duration of our sequence. Because sapphire's waves are only resolvable near the focus, our images should show them as stationary rings that change in intensity between frames.

The depth-variation in 3D sapphire waves at the focus highlights a further point that is a consequence of the imaging lens, shown in Figure 7. The ~4-μm depth of focus along the $Z$-axis, set by the 10X objective, places the substrates and air outside of the image plane ($Z_i$). This means that some of the water layer produces a shadowgraph image at the detector, while the remainder of the target generates superimposed phase-sensitive Talbot images.[33] Changes to the index of refraction in each $R_S$-$\theta$ plane along $Z$ (through all layers) can generate additional features in these images that originate from overlapping Talbot images (Fig. 7). These effects are small for most of the target—which see very small $\Delta n$ variation—but at the center of convergence, the sapphire reaches a sufficiently high pressures to produce Talbot signal. As indicated in Figure 7, we attribute those image features to the converging waves in the sapphire layers.

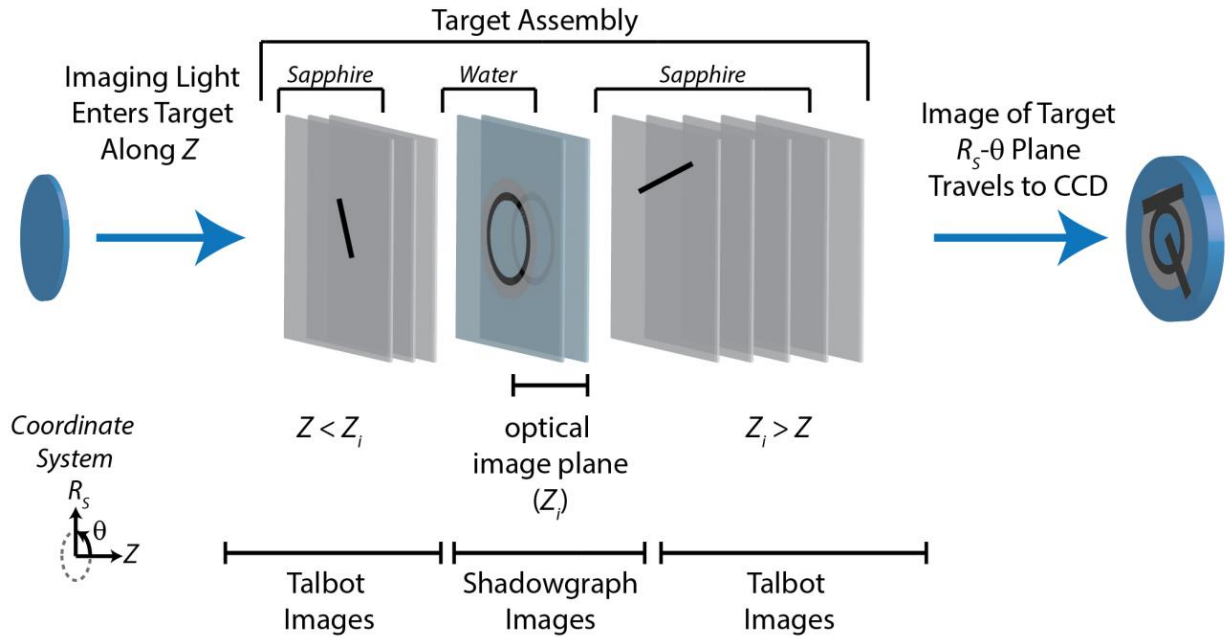

**Figure 7.** Illustration of signal contributions in our shadowgraph imaging system from different regions along the Z-axis of our target assembly. Shadowgraph and Talbot contributions from different target layers are overlaid in the images.

Qualitative comparison of our dark-field and shadowgraph image sequences show that the different image sources provide complementary information about the shock behavior. However, shot-to-shot fluctuations in the shock positions and velocities prevent quantitative numerical comparisons between the features produced between these image sequences. While the converging shock geometry amplifies the shock pressure, it also amplifies the ~5% energy fluctuations in our drive pulse, creating geometric instabilities.[20] This shot-to-shot variation prevented previous experiments from tracking the temporal progression of the additional waves,[6] as is done here. While comparison between different sequences can only be qualitative or statistical, individual frames within a single sequence may be compared quantitatively. This technique's ability to track irreproducible phenomena also enables it to monitor the temporal progression of fracture and deformation of shock geometries, which can be valuable for future experiments.

The combination of the dark-field and shadowgraph imaging sequences have given us a new and more detailed view of shock waves in our quasi-confined geometry. In both cases, the unanticipated extra rings in our images could be interpreted because the single-shot sequences enabled us to track the motion of each ring in our images. Our dark-field images gave a set of rings in each image that showed relatively subtle density changes for the oblique front and release, but over-rotation created extra features within the primary shock. This information was complemented by the shadowgraph image sequence which contained mostly information about the primary water wave. The different sensitivities of the two methods to different features of the shock enabled us to interpret each of the additional rings that we observed in our shadowgraph image sequence, most of which originated from the water components of the coupled wave caused by the interface.

This more detailed view of our setup gives us additional perspective on shock waves in this quasi-confined experimental geometry. Firstly, material studies using this geometry require carefully designed targets, for which the substrate shocks do not overwhelm the signal from the sample layer. For imaging experiments like the ones presented here, this requires small substrate photoelastic constants.

Secondly, the coupled wave structure shows the importance of modeling target geometries with materials of different impedances. As the impedance mismatch between the sample and substrate layers sets the relevant speeds for our coupled-wave structure, tuning the relative impedances can adjust the wave interactions. For the study of solid sample materials, inverting the impedance mismatch to use a high-impedance sample layer between two low-impedance substrates could simplify the coupled-wave structure since in that case the fastest wave speed would be in the substrate. We have used dye-doped, low-impedance polymers for absorption of the laser light and generation of shocks, and these might be useful for the outer substrate layers. This approach is not possible for liquid samples, but thicker liquid layers can help reduce the effects of interfaces with the substrates. Further work is ongoing to explore modified sample configurations.

The coupled wave geometry introduces complexity in determination of the *P-T* state of the system. Previously, we inferred the average shock pressure by inputting the measured shock velocity between frames to the principal Hugoniot equation of state.[6,12] The principal Hugoniot relates the shock velocity to the thermodynamic variables that describe the material state just behind the shock front, assuming the shock encounters a material initially at rest. In this system, the leading oblique waves in the coupled-wave pre-shock the material before the primary water wave arrives, invalidating this assumption. Details of the pre-shock are discussed in the Supplemental Information with logarithmic pressure plots that allow moderate pressures to be highlighted.

Further studies are required to provide more precise measurement of sample thickness and shock speed to allow for the thermodynamic conditions of the sample to be quantitatively inferred.

**Conclusions**

This paper presents the development of single-shot multi-frame imaging that extends our understanding of a tabletop platform for generation and measurement of quasi-2D converging shock waves.[6] We used a frequency-doubling Fabry-Perot cavity to generate a train of femtosecond pulses for illumination of all the frames. This experimental scheme, which can incorporate a range of different imaging methods, is shown here for shadowgraph and dark-field imaging, using a high-frame-rate camera. In our multi-layered target geometry, the image sequences all showed multiple concentric rings originating from the primary shock in the water sample layer and from additional shock structure. The coupled-wave structure caused by the sapphire-water interfaces caused additional oblique waves in the water that created additional rings in our image sequences. The complex impedance-dependent coupled-wave structure shows effects that expand our previous understanding of the shocks within this target assembly. Our imaging technique and understanding of the coupled-wave structure can inform future material studies in this tabletop experimental geometry. The single-shot imaging presented here is useful in this and other experimental geometries for understanding shock behavior with temporally complex dynamics with shot-to-shot variation. The results also suggest experimental steps that could simplify the shock behavior for solid samples, including an inverted geometry in which the fastest shock is the primary wave within the (high-impedance) sample of interest.

The tabletop technique we present here provides an easily implementable shock experiment that may be used to study how materials respond to non-uniaxial shock waves. As the experiment only uses small volumes of the sample, it can be used as a high-throughput measurement tool for optimizing material designs for specific functionality during shock waves. We have demonstrated how the tabletop experiment allows us to design new diagnostic tools, which may be used in conjunction with imaging. The experimental design with the probe light arriving at the sample perpendicular to the shock propagation spatially separates the different regions of the shock wave, which can enable spectroscopy to resolve how the material response evolves during the shock. The opportunities for multiple simultaneous *in-situ* probes such as multiframe imaging and future spectroscopy in this experiment presents a unique toolbox for studying how materials change during shock waves.

## Acknowledgements

The first author would like to commemorate and thank Professor Mildred S. Dresselhaus for her support, advice, guidance and mentorship in all of this and other work. The authors thank Michelle Rhodes for her guidance and assistance in developing optical simulation code that informed some conclusions about photoelasticity. The authors would like to thank Marylesa Howard for her assistance in the analysis of the images sequences collected in this work. We further acknowledge Emma McBride and Siegfried Glenzer for their discussions and support in this work. The work at MIT was supported in part by the Office of Naval Research grants N00014-16-1-2090, N00014-15-1-2694, and (DURIP) N00014-15-1-2879.

# Supplemental Information

## Image Processing Description

The image sequences shown in this work were processed using the white balance, contrast and noise enhancement features from ImageJ. The images .TIF images collected from the camera were first manually white balanced, then noise filters were applied. We used the built-in despeckle to remove the random pixel noise produced by gain in the detectors. Figure S-1 shows the entire shadowgraph image sequence before (a) and after (b) the despeckle filter was applied. The image processing did not change the visible or resolvable image features, in all cases.

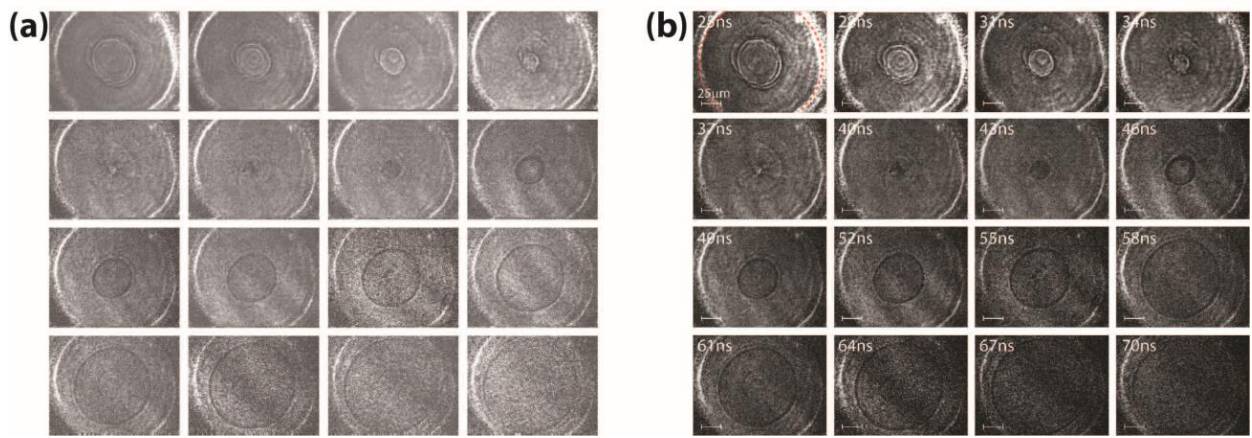

**Figure S-1.** Image sequence from Figure 1 in the main paper (a) after the white-balance but before the contrast enhancement and subsequent despeckle filtering, and (b) processed.

## Description of Simulations

Our simulations included all five layers in the target. No assumptions were made regarding local boundary conditions, allowing interface locations to vary. Computational domain boundaries were placed far enough from the experimental region of interest that artificial boundary wave interactions were avoided during the time scale of the simulated experiments.

The simulations do not treat laser light-matter interactions explicitly. We assume that 90% of the energy from the drive laser pulse is transferred into the water (based on previously reported sample transmission[1]; no light absorption in other target layers is considered) with spatial and temporal profiles based on the experimental parameters, i.e. 150 μm laser ring inner diameter, 8 μm thick laser ring line width, and 100 ps Gaussian temporal profile. Rapid heating of the irradiated region induces a compressive shock, which is alleviated by a rapid expansion that initiates bulk shocks in the target (and the subsequent shock dynamics). The laser-induced shock in the simulation is symmetric about the plane of the water layer, such that the shock pressures in the figure are symmetric above and below the Z axis (centered in the middle of the water layer).

Simulations for the wave dynamics were performed using the CTH shock physics code, developed by Sandia National Laboratories.[2] A Mie-Grüneisen equation of state with the Hugoniot as reference was used to describe the hydrodynamic and thermodynamic behavior of the water sample. The principal Hugoniot was determined from a quadratic fit to shock velocity-particle velocity data with parameters $c_0$ = 1.48 km/s, $s_1$ = 1.984, $s_2$ = -0.143, a reference density $\rho_0$ = 0.998 g/cm$^3$, Grüneisen parameter, $\Gamma_0$ = 0.48, and a specific heat, $C_V$ = 3.69 J/g/K.[3] Sesame tabular equation of state and elastic-perfectly plastic strength models described the behavior of the sapphire substrates. The outer air layers were also modeled with a Sesame equation of state. The computations were performed in an axisymmetric geometry, using a uniform mesh size of 0.25 μm throughout the region of interest.

## Details of the Pre-Shock

A detailed view of the pre-shock is shown in Figure S-2, which shows three time slices from our simulated results as the primary water shock converges with a logarithmic pressure scale. This shows that the pre-shock develops and increases in intensity as the primary water wave converges, with a highly non-uniform structure across the Z-axis. At 1 ns, as the shocks begin to travel, no pre-shock is evident in the water layer. As the water and sapphire shocks begin to travel, the pre-shock begins to form, as seen by 7 ns. In Figure S-2b, we see a significant pre-shock has formed in the water. The surface component of the initial 3D sapphire wave creates a weak oblique water wave of 10$^{-3}$ GPa, as the sapphire component of the coupled-wave structure begins to move ahead of the primary water wave. At this time, the oblique sapphire waves and primary water shock have not separated enough to induce an additional oblique water wave in the coupled-wave structure. Over the subsequent 10 ns, the oblique sapphire wave speeds ahead of the primary water wave, and leaks further energy back into the water layer. By 16 ns, the secondary oblique water wave has a steeply angled front making it nonuniform along the Z-axis, and includes a range of 0.01-1 GPa in pressure.

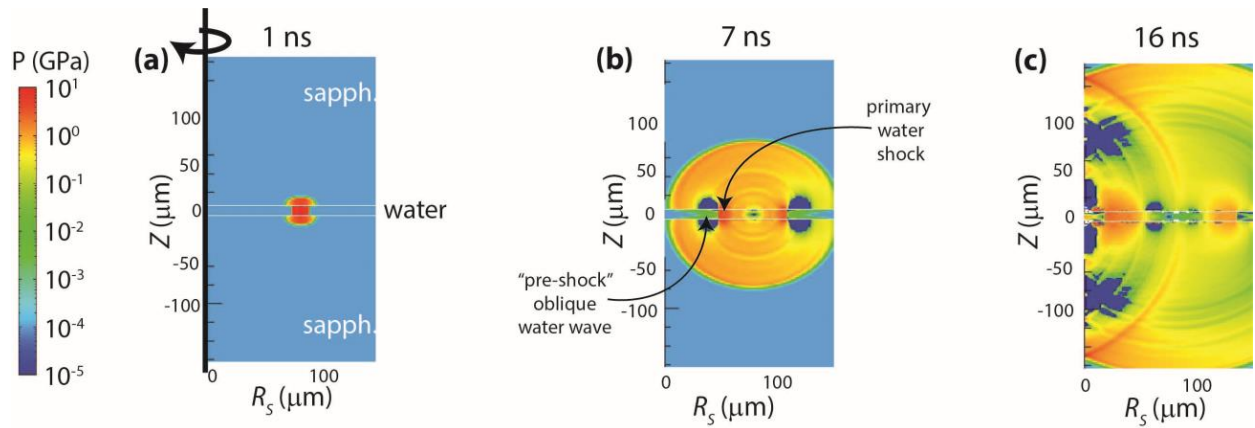

**Figure S-2.** Simulated results showing a logarithmic pressure scale to emphasize the low pressure disturbances in the water layer.

Through its entire propagation, the pre-shock only ever reaches a maximum pressure of ~1 GPa when at the center of convergence. During its propagation, the pre-shock typically shows a pressure between 0.01 and 0.5 GPa, whereas the sapphire waves are ~7 GPa upon convergence while the primary water shock stays around ~10 GPa. At almost all times, the pre-shock has at least an order of magnitude lower pressure than the primary water wave. Quantitative studies are required to understand how much the pre-shock shifts the *P-T* conditions in the primary water wave off the principal Hugoniot.